\documentclass[11pt,twoside]{article}

\usepackage{asp2014}

\aspSuppressVolSlug
\resetcounters

\begin{document}

\title{SIPGI: an interactive pipeline for spectroscopic data reduction}

\author{Susanna Bisogni,$^1$ Adriana Gargiulo,$^1$ Marco Fumana,$^1$ Paolo Franzetti,$^1$ Letizia P. Cassarà,$^1$ Marco Scodeggio,$^1$ Bianca Garilli $^1$ and Giustina Vietri$^2$}
\affil{$^1$INAF-IASF Milano, via Alfonso Corti 12, 20133 Milano, Italia; $^2$Osservatorio Astronomico di Roma (INAF), via Frascati 33, 00040 Monte Porzio Catone (Roma), Italy;
\email{susanna.bisogni@inaf.it}}

\paperauthor{Susanna Bisogni}{susanna.bisogni@inaf.it}{ORCID_Or_Blank}{Author1 Institution}{Author1 Department}{City}{State/Province}{Postal Code}{Country}
\paperauthor{Sample~Author2}{Author2Email@email.edu}{ORCID_Or_Blank}{Author2 Institution}{Author2 Department}{City}{State/Province}{Postal Code}{Country}
\paperauthor{Sample~Author3}{Author3Email@email.edu}{ORCID_Or_Blank}{Author3 Institution}{Author3 Department}{City}{State/Province}{Postal Code}{Country}
\paperauthor{Sample~Author3}{Author3Email@email.edu}{ORCID_Or_Blank}{Author3 Institution}{Author3 Department}{City}{State/Province}{Postal Code}{Country}


\begin{abstract}

SIPGI is a spectroscopic pipeline for the data reduction of optical/near-infrared data acquired by slit-based spectrographs. SIPGI is a complete spectroscopic data reduction environment retaining the high level of flexibility and accuracy typical of the standard “by-hand” reduction methods but with a significantly higher level of efficiency. This is obtained exploiting three main concepts: 1) a built-in data organiser to classify the data, together with a graphical interface; 2) the instrument model (analytic description of the main calibration relations); 3) the design and flexibility of the reduction recipes: the number of tasks required to perform a complete reduction is minimized, preserving the possibility to verify the accuracy of the main stages of data-reduction process. 
The current version of SIPGI manages data from the MODS and LUCI spectrographs mounted at the Large Binocular Telescope (LBT) with the idea to extend SIPGI to support other through-slit spectrographs.

\end{abstract}

\section{Introduction}

SIPGI \citep{Gargiulo2022} is a complete spectroscopic reduction environment for optical and near-IR through-slit spectra developed for the astronomical community and now publicly released. 
Data reduction is usually a time-consuming process; this pipeline was designed to perform a quick data reduction, while mantaining a high level of flexibility and accuracy.
SIPGI works with both optical and near-IR data. It features a graphical interface and many tools that strongly improve the data reduction experience.

The version released to the community is customised for the data acquired with the two couples of through-slit spectrographs @LBT, MODS1 and MODS2 in the optical band and LUCI1 and LUCI2 in the near-IR. However, the code minimises the dependence from the specific spectrograph, confining the information on the instrument to specific parts of the code; this makes the pipeline adaptable in the future to any through-slit (and possibly fibers) spectrograph.

\section{The main SIPGI concepts}

The high efficiency and flexibility can be simultaneously achieved by exploiting three concepts: the Graphical Interface, the Instrument Model and the Recipes organisation.
SIPGI inherits them from VIPGI \citep{Scodeggio2005}, the pipeline we developed and used for the data reduction of the main extragalactic spectroscopic surveys carried out with the optical spectrograph VIMOS@VLT (e.g. VVDS, zCOSMOS, VUDS, VIPERS, VANDELS).

\subsection{The Graphical Interface and the Data Organiser}

\articlefigure[width=1.0\textwidth]{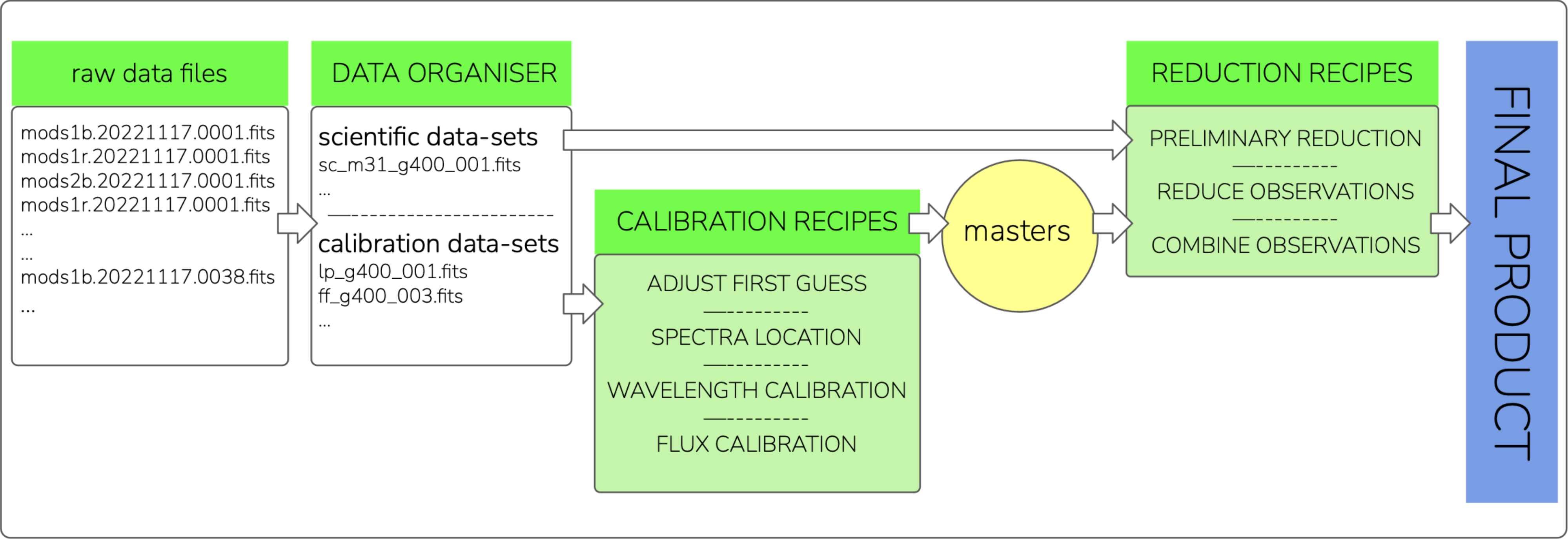}{label_flowchart}{\footnotesize{SIPGI flowchart.}}

SIPGI ingests all the raw frames needed for the data reduction (calibration and scientific files) and, thanks to the Data Organiser, it recognises and classifies them according to the keywords in their header (Fig. \ref{label_flowchart}).
All the data are grouped in data-sets (Fig. \ref{labelGraphInt} A), collections of data with the same characteristics. The user can easily browse through different data-sets, divided in calibration (acquired through/without slit) and scientific data-sets (target and standard star) and, within each data-set, through the different reduction units (Fig. \ref{labelGraphInt} B), defined by the instrumental configuration they have been observed with, i.e. the mask, camera, grating and dichroic used during the observations.

\articlefigure[width=0.95\textwidth]{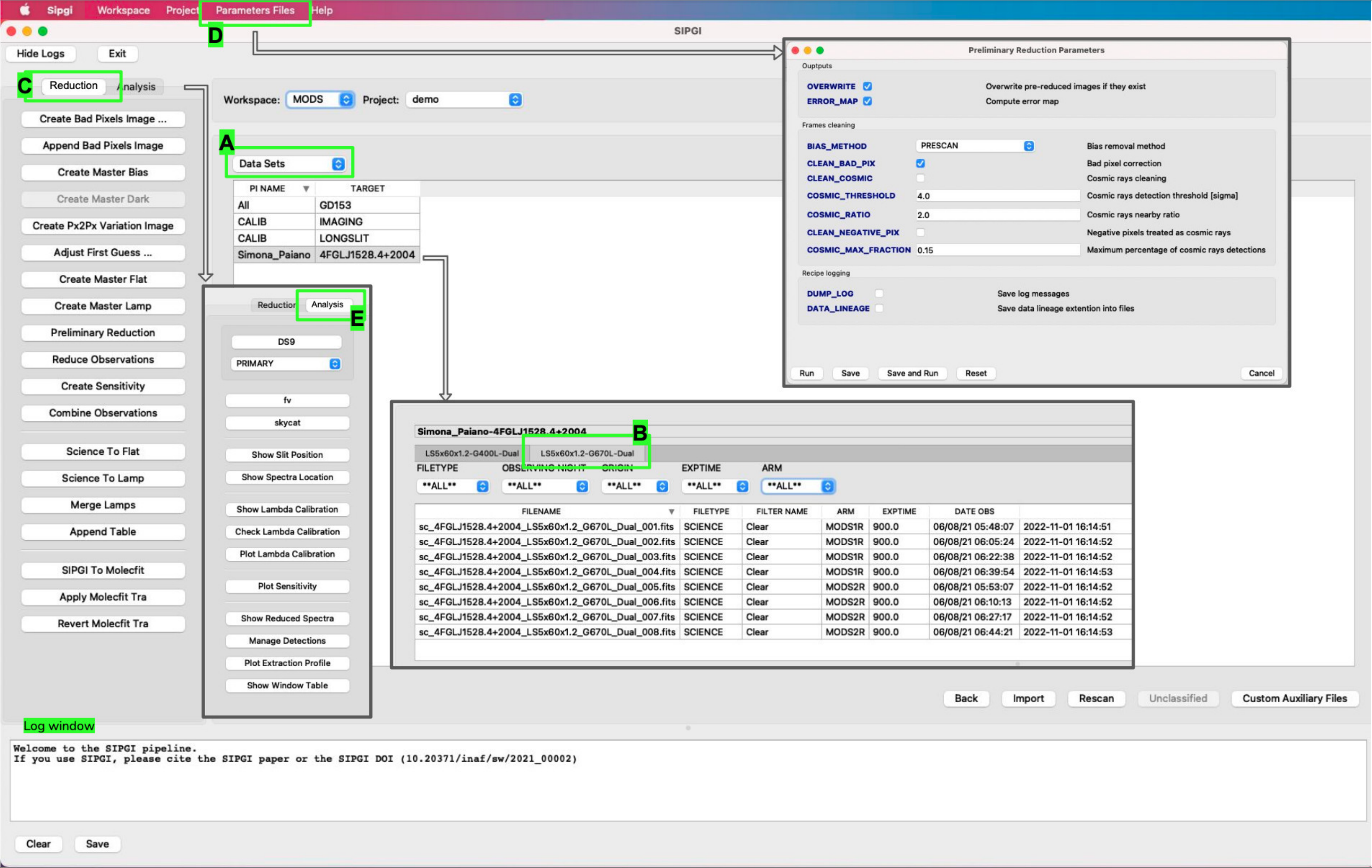}{labelGraphInt}{\footnotesize{SIPGI Graphical Interface. 
Data are organised in data-sets (A), containing one or more reduction units (B). The recipes can be run from Reduction Tab (C) and their parameters files can be accessed and modified by the user (D, example for the Preliminary Reduction). 
The Analysis Tab (E) offers tools and utilities for checking mid and final products.
The frames are renamed to be easily recognisable (e.g. "sc" for scientific, "lp" for lamp, "ff" for flat etc.).}}

The graphical interface allows to run the recipes by pressing the buttons displayed in the Reduction Tab (Fig. \ref{labelGraphInt} C). 
The Parameter Files menu gives a quick access to the parameters file for each recipe (Fig. \ref{labelGraphInt} D, see section \ref{par_recipes}).
The Analysis Tab (Fig. \ref{labelGraphInt} E) features several graphical tools to check the quality of the raw frames, mid-products and final products of the data reduction and specifically designed utilities to verify the accuracy of the calibrator files produced during the data reduction.

\subsection{The Instrument model}

The Instrument Model is an analytical description of the main calibration
relations necessary to obtain rectified spectra from the
observations performed with a spectrograph. 
It depends only on the instrument configuration (i.e. grating/filter/dichroic) and is therefore mask-independent. It has been calibrated on real data and it is provided with SIPGI.
It provides the first guesses to the spectra location (the position of the slit edges in the frames) and to the wavelength calibration.
Since the distortions in the frames can change on a night basis, the first guesses need to be updated on the specific set of data. This is done through the use of the Adjust First Guess tool on a set of calibration frames specifically acquired for the program: the user can visualise the model, represented by a grid of ds9 regions, on a real lamp frame (Fig. \ref{label_Adjust}) and find the best match between the real data and the first guesses of the spectra location (vertical lines) and wavelength calibration (horizontal lines).

The concept of the instrument model is what really lightens the task of calibrating the data. 
If the instrument is stable, the first guesses are already an excellent approximation of the real distortions in the frame and the user needs to provide only small, if any, adjustments. Even in case of instability, the graphical tool makes it much less painful to find an agreement between the analytical description and the real distortions.
Furthermore, the mask-independence of the model is a key-point in terms of time-saving: when multiple masks are being used in an observing program, the same calibrators produced for one of them can be, in principle, reused for the others.
\articlefigure[width=0.5\textwidth]{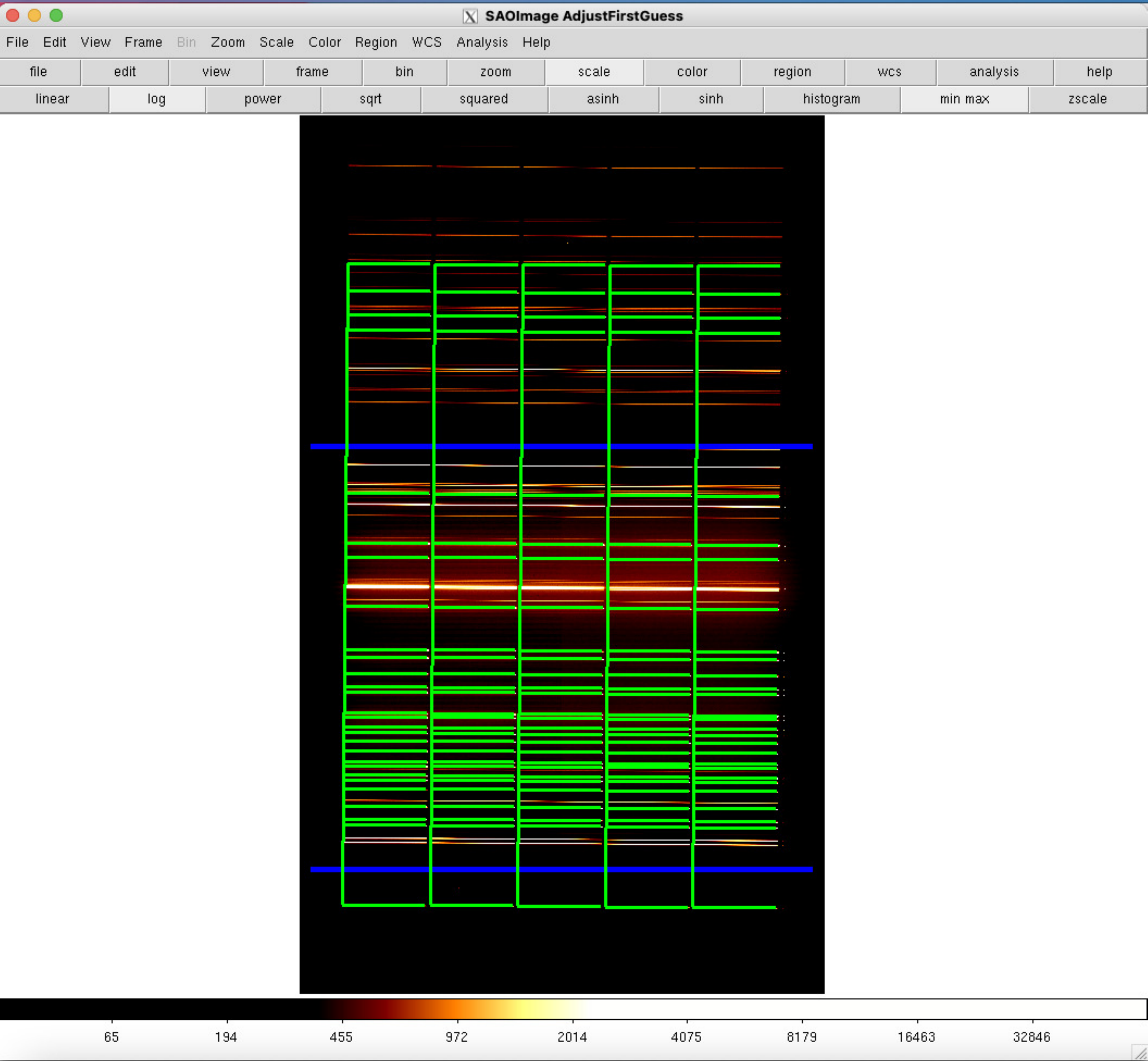}{label_Adjust}{\footnotesize{Adjust First Guess tool. The graphical representation of the Instrument Model, i.e. of the first guesses to the spectra location (vertical lines) and wavelength calibration (horizontal lines), is superimposed to a lamp frame of the data set to be reduced.}}

\subsection{The recipes organisation} \label{par_recipes}

The efficiency of SIPGI is achieved through the design of the recipes.
Each recipe performs many tasks, reducing the time needed for completing the data reduction.
Four calibration recipes produce the master frames used to infer spectra location, wavelength and flux calibration, while three reduction recipes apply the solutions provided by the master frames to the scientific frames (Fig. \ref{label_flowchart}).
Each recipe has many parameters that can be set through the Parameter Files Menu (Fig. \ref{labelGraphInt} D). This allows the user to highly customised their data reduction to meet their scientific goals.

We decided to separate the reduction recipes in three different steps (see Fig. \ref{label_flowchart}): 1. a first one removing the instrument signature, performing a bias (or dark) level subtraction, the flat-fielding and the cosmic rays cleaning (Preliminary Reduction); 2. the main step of the reduction, producing the 2D and 1D rectified, wavelength and flux calibrated spectra for individual frames (Reduce Observations); 3. the recipe combining individual frames in a final product (Combine Observations). This architecture allows the user to obtain mid-products, that can be checked through many tools expressly designed within SIPGI and, possibly, treated with customised routines if desired.

\section{Software architecture and performances}

SIPGI provides a GUI written in Python, used to organise data, to run the reduction recipes (written in C) and to check reduction results with a set of interactive graphical tools. The interaction between Python and C is obtained using the SWIG wrapper.

The reliability of SIPGI is attested by its extensive use for the data reduction of all the data acquired with LUCI and MODS at LBT, during the italian time, in the last ten years, whose products were used in $\sim$80 referred publications. 
For all the standard configurations with both the instruments, SIPGI provides wavelength calibration with an accuracy better than 1/5 of pixel in 95 per cent of the cases.
The typical rms in the flux calibration is 0.4\% (0.5\%) in
the regions not affected by telluric absorption for MODS (LUCI) data
and 2\% (5\%) in regions affected by telluric absorptions.

SIPGI execution performances obviously depend on the computer hardware exploited and on the kind of data being reduced. However, for a standard 3h-observations set of data in binocular mode (i.e. MODS1+MODS2 or LUCI1+LUCI2) we estimate an average execution time of three hours.

\vspace{0.3cm}

Further information on SIPGI, the download page, a manual and cookbook, and a series of video tutorial can be found at \url{http://pandora.lambrate.inaf.it/sipgi/}.
The users can contact us at the help-desk \email{lbt-italia-spec@inaf.it}.

\bibliography{F03}  


\end{document}